\documentclass[rnote]{aa}
\usepackage{graphicx}

\usepackage{txfonts}

\begin{document}

\title {Chemical evolution of the Milky Way: the origin of phosphorus}

\author {G. Cescutti\inst{1} \thanks {email to:
    gabriele.cescutti@epfl.ch}\and F. Matteucci\inst{2,3}\and
  E.Caffau\inst{4,5}\and P. Fran\c cois\inst{5,6}} 
\institute {Laboratoire d'Astrophysique, 
 Ecole Polytechnique F\'ed\'erale de Lausanne (EPFL),
 Observatoire de Sauverny,
 CH-1290 Versoix, Switzerland
\and Dipartimento di Fisica, Sezione di Astronomia, 
Universit\`a di Trieste, via
  G.B. Tiepolo 11, I-34131, Trieste, Italy 
\and I.N.A.F. Osservatorio
  Astronomico di Trieste, via G.B. Tiepolo 11, I-34131, Trieste,
  Italy
\and Zentrum f\"ur Astronomie der Universit\"at Heidelberg, Landessternwarte, 
K\"onig stuhl 12, 69117 Heidelberg, Germany
\and GEPI, Observatoire de Paris, CNRS Universit\'e Paris Diderot, Place 
Jules Janssen, 92190, Meudon, France
\and Universit\'e de Picardie Jules Verne, 33 Rue Saint Leu,  Amiens, France
}

\date{Received xxxx / Accepted xxxx}

\abstract {Recently, for the first time the abundance of P has been
  measured in disk stars. This provides the opportunity of comparing
  the observed abundances with predictions from theoretical models.}
{We aim at predicting the chemical evolution of P in the Milky Way and
  compare our results with the observed P abundances in disk stars in
  order to put constraints on the P nucleosynthesis.}  {To do that we
  adopt the two-infall model of galactic chemical evolution, which is
  a good model for the Milky Way, and compute the evolution of the
  abundances of P and Fe. We adopt stellar yields for these
  elements from different sources. The element P should have been
  formed mainly in Type II supernovae. Finally, Fe is
  mainly produced by Type Ia supernovae.}  {Our results confirm that
  to reproduce the observed trend of [P/Fe] vs. [Fe/H] in disk stars,
  P is formed mainly in  massive stars. However, none of
  the available yields for P can reproduce the solar abundance of this
  element.  In other words, to reproduce the data one should assume
  that massive stars produce more P than predicted by a factor of
  $\sim$ 3. } {We conclude that all the available yields of P from
  massive stars are largely underestimated and that nucleosynthesis 
 calculations should be revised. We also predict the [P/Fe] expected 
in halo stars.
  }


\titlerunning{The origin of phosphorus}
\authorrunning{Cescutti et al.}
\maketitle
 
\section{Introduction}
Recently, Caffau et al. (2011a) have measured the P abundance in a
sample of 20 cool stars in the Galactic disk. They found that the
[P/Fe] ratio behaves like the [S/Fe] ratio, namely increasing towards
lower metallicity stars. This was the first time that P was observed
in Galactic stars in spite of the fact that P is among the top 20 most
abundant elements in the Universe.  There is only one single stable
isotope of P, $^{31}$P, which is thought to be formed by neutron
capture on $^{29}$Si and $^{30}$Si in massive stars. By means of a
detailed model for the chemical evolution of the Milky Way one can
compute the P evolution and compare it to the observations. This can
allow us to understand the origin of this element and impose
constraints on its formation inside stars.  In this paper, we adopt a
good model for the chemical evolution on the Galaxy, already tested on
a large number of chemical species (Fran\c cois et al. 2004). In
particular, the model takes into account detailed stellar
nucleosynthesis and supernova progenitors (SNe Ia, II, Ib/c) as well
as the stellar lifetimes.  The stellar yields of P have been computed
by several authors such as Woosley \& Weaver (1995), Kobayashi et
al. (2006) as functions of the initial stellar metallicity.  In this
paper we like to compare the predictions of our chemical evolution
model for P, obtained by including different stellar yields, with the
recent data.  In Section 2 we briefly describe the chemical evolution
model adopted, in Section 3 we describe the observational data. In
Section 4 the results are compared to the data and finally, in Section
5, some conclusions are drawn.

\section{The chemical evolution model}
The role played by SNe of different Type in the chemical 
evolution of the Galaxy has been computed by several authors in the past years
(Matteucci \& Greggio, 
1986; Matteucci \& Fran\c cois 1989; Yoshii et al. 1996;  Chiappini et al. 1997, 2001; Kobayashi etal. 2006; Boissier \& Prantzos 1999; Fran\c cois  et al . 2004 among others).
Here we refer to the model of Chiappini et al. (1997), the so-called two-infall model for the 
evolution of the Milky Way. A thorough description of this model can be found in Chiappini et al. 
(1997; 2001) and Fran\c cois et al. (2004) and we address the reader to these 
papers for details. 
In this model it is assumed that the stellar halo formed on a 
relatively short 
timescale (1-2 Gyr) by means of a first infall episode, whereas the disk 
formed much more slowly, mainly out of 
extragalactic gas, thanks to a second infall episode. The timescale for the 
disk formation is assumed to increase 
with the galactocentric distance ($\tau= 7$ Gyr at the solar circle), 
thus producing an ``inside-out'' scenario 
for the disk formation. The Galactic disk is divided in several rings, 2Kpc 
wide, without exchange of matter between them.  
The model can follow in detail the evolution of several chemical 
elements including H, D, He, C, N, O, $\alpha$-elements, Fe and Fe-peak elements,  s- and r-process elements.
The star formation rate (SFR) adopted for the Milky Way is a function of both 
surface gas density and total surface mass density.
Such a SFR is proportional to a power $k = 1.5$ of the
surface gas density and to a power $h = 0.5$ of the total surface mass
density. 
In the Milky Way model 
we also assume a surface density threshold below which the SFR stops, 
according to Kennicutt (1989). As a consequence of
this, the star formation rate goes to zero every time that the gas
density decreases below the threshold ($\sim 7
M_\odot\,pc^{-2}$).
The efficiency of SF is $\nu=1 Gyr^{-1}$ during the disk 
and 2$Gyr^{-1}$ in the halo phase. 
This  model reproduces the majority of the features
of the solar vicinity and the whole disk.

\subsection{Nucleosynthesis and stellar evolution prescriptions}
We have adopted the yields for P and Fe originating in massive stars
from Woosley \& Weaver (1995, hereafter WW95) and Kobayashi et
al. (2006, hereafter K06) as functions of stellar metallicity.  The
yields for the same elements originating from SNe Ia are from Iwamoto
et al. (1999), their model W7.  In particular, each SN Ia is assumed
to eject the same mass, the Chandrasekhar mass ($\sim 1.4 M_{\odot}$); of
this mass $\sim 0.6 M_{\odot}$ is in the form of Fe and $\sim 3.57
\cdot 10^{-4}$ are in the form of P. It is therefore clear that SNe Ia
are negligible producers of P. On the other hand, $^{31}$P should be produced 
during O- and Ne- shell burnings in massive stars, although a large 
fraction of $^{31}$P can be destroyed by (p,$\alpha$) reactions to 
become $^{28}$Si.

\section{The observational data}

The sample of observed stars is formed by 20 G-F bright dwarfs
($3.3\le{\rm J_{\rm mag}} < 7$),
with  stellar parameters in the ranges:
$5765\le {\rm T_{\rm eff}}\le 6470\,$ K,
$3.90\le {log g} < 4.5$,
$-0.91\le\left[{\rm Fe/H}\right]\le 0.28$.

The phosphorus abundance has been derived from CRIRES spectra
observed at VLT-Antu 8\,m telescope, in service mode during ESO period
86. The setting was centered at
1059.6$\,$ nm in order 54. Four P lines of Mult.$\,$1
are visible in detectors 2 and 3.
The spectral resolution was R=100$\,$000,
the signal-to-noise ratios achieved were in the range 50-400.
For the Adaptive Optics correction, computed on axis,
the target star itself was used.

\section{Results for phosphorous} 

We have run several models including different prescriptions for the P and Fe
yields from massive stars. 
In some of the models we mixed yields from different sources and this is 
allowed by the fact that yields of some specific elements, in particular 
those of P 
and Fe, are still uncertain, because of the different input physics adopted by 
different authors. As a consequence, by means of our chemical evolution 
models we can put constraints on the nucleosynthesis calculations.

The models are:

\begin{itemize}
\item Model 1: with the P yields from massive stars of K06
  and the Fe yields from massive stars of WW95 considering only the
  solar chemical composition. This choice is dictated by the fact that
  these Fe yields better reproduce the [X/Fe] vs. [Fe/H] relations
  (see Fran\c cois et al. 2004);
\item Model 2: with the P yields from massive stars of K06
  and metallicity dependent Fe yields from massive stars of WW95;
\item Model 3: with the metallicity dependent P yields from massive
  stars of WW95 and the Fe yields from massive stars of WW95
  considering only the solar chemical composition;
\item Model 4: with metallicity dependent P and Fe yields
  from massive stars of WW95;
\item Model 5: with P and Fe metallicity dependent yields
  from massive stars of K06;
\item Model 6: where the yields of P and Fe from massive
  stars are from K06 and include also hypernovae (hypernovae are stars
  with $M> 20M_{\odot}$ and explosion energies larger than normal SNe II); 
\item Model 7: the only difference compared to model 5 is that 
in this model the P yields in massive stars are artificially increased by a 
factor of $\sim$ 3;
\item Model 8: the only difference compared to model 6 is that 
in this model the P yields in massive stars are artificially increased by a 
factor of $\sim$ 2.75.
\end{itemize}

In order to obtain the [P/Fe] ratios, we have normalized our model
results for P and Fe with the observed absolute solar abundances
obtained by Caffau et al. (2011b).
In this way, one can see at the same time 
whether the model predictions fit the trend as well as the solar abundances.
The predicted solar abundances are the abundances that the model predicts for 
the ISM 4.5 Gyr ago, the time of formation of the solar system.

In Table 1 we show a
summary of the nucleosynthesis prescriptions used for P and Fe in the models 
and the difference between the predicted solar abundances and the observed ones
by Caffau et al. (2011b).
Note that the yields of P and Fe from SNe Ia are from Iwamoto et
al. (1999) in all models.

\begin{table*}
\begin{center}
\caption{Nucleosynthesis prescriptions for the various models.}
\begin{tabular}{|c|c|c|c|}
 \hline
  Model    &    P from massive stars  &   Fe from massive stars  & [P/Fe]$_{\odot}$  model    \\  
 \hline
   1      &      K06        &    WW95 solar chem. comp.  & -0.41\\
\hline
   2       &   K06     &  WW95 met.dep.  & -0.42 \\
\hline
   3      &    WW95 met. dep.   & WW95 solar chem comp. & +0.06 \\
 \hline
   4     &   WW95 met. dep.   &  WW95 met.dep.  & +0.05\\
\hline
    5       &    K06      & K06    & -0.40\\
\hline 
    6     &     K06+ Hypern. &  K06+ Hypern & -0.37\\
\hline
    7       &  3 $\times$  K06      & K06    & 0.00  \\
\hline 
    8     &    2.75 $\times$ K06 + Hypern. &  K06+ Hypern. & 0.00 \\
\hline
\end{tabular}
\end{center}
\
\label{tabmodels}
\end{table*}

\begin{figure}[htb]  
    \begin{center}  
    \includegraphics[width=0.48\textwidth]{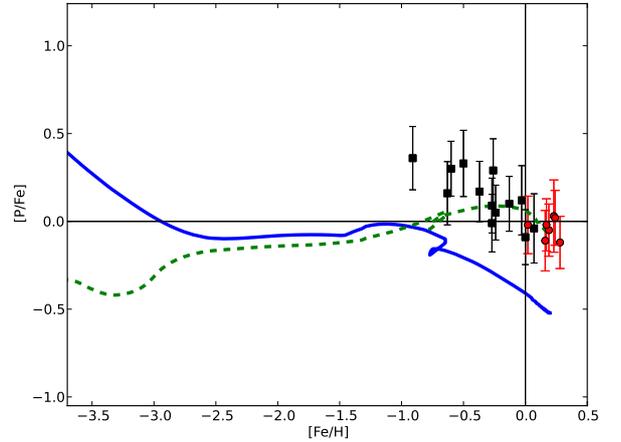}
    \caption{The predictions from Model 1 (blue solid line) and from Model 3
      (green dashed line). The P and Fe abundances in both models are normalized with the absolute solar values of Caffau et al. (2011b).
      In Table 1 are shown the predicted ratios at the age of formation of the solar system. Data from Caffau et al. (2011a): the circles indicate stars with planets.} 
    \label{Fig01} 
    \end{center}
    \end{figure}

In Figure 1 are shown the results of Models 1 and 3. 
As one can see, for Model 1 the agreement with the
observed trend in the Galactic disk is good, although the curve is shifted to lower values than observed. This is due to the fact that while the solar Fe
abundance predicted by this model is in good agreement with the observed one, the P solar abundance is too low. In particular, in order to reproduce the P solar abundance 
we need to increase artificially the yields of P from massive stars by a 
factor of $\sim$ 3.  
On the other hand, we note that the results of Model 3
 are not able to
reproduce the trend of the [P/Fe] ratio in the Galactic disk although the predicted solar value is almost acceptable (+0.06 dex, in Table 1).

\begin{figure}[htb]  
    \begin{center}  
    \includegraphics[width=0.48\textwidth]{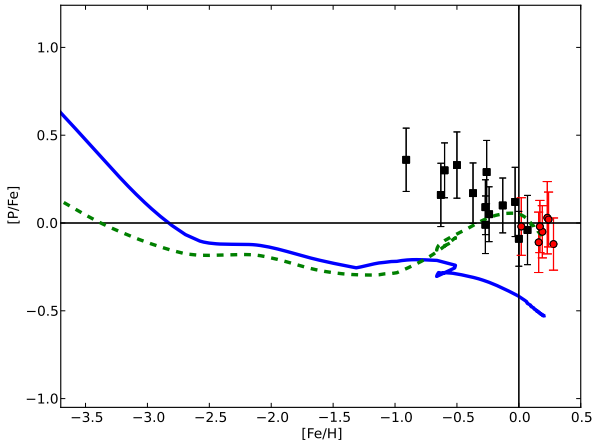}
    \caption{The predictions from Model 2 (blue solid line) and from Model 4
      (green dashed line) . The P and Fe abundances in both models are normalized with the absolute solar values of Caffau et al. (2011b). 
     In Table 1 are shown the predicted ratios at the age of formation of the solar system. The data are the same as in Figure 1.}
    \label{Fig02} 
    \end{center}
    \end{figure}

In Fig. \ref{Fig02}  are  shown Models 2 and 4: for Model 2 the agreement with the observed trend in the Galactic disk is good, whereas for Model 4 the trend is not reproduced, although, as for Model 3, the predicted solar value could be acceptable (+0.05, Table 1). 
To obtain a good agreement with the observed solar abundance of P for 
Model 2, we need again to increase 
artificially the
yields of P by a factor of $\sim$ 3.

\begin{figure}[htb]  
    \begin{center}  
    \includegraphics[width=0.48\textwidth]{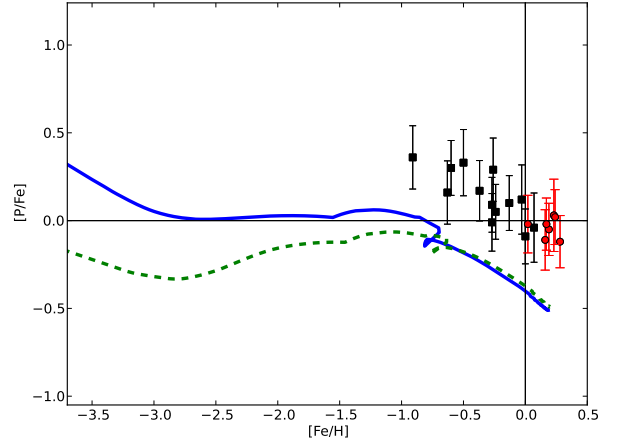}
    \caption{The predictions from Model 5 (blue solid line) and from Model 6
      (green dashed line). The P and Fe abundances in both models are normalized with the absolute solar values of Caffau et al. (2011b). 
      In Table 1 are shown the predicted ratios at the age of formation of the solar system. The data are the same as in Figure 1.} 
    \label{Fig03} 
    \end{center}
    \end{figure}

In Figure \ref{Fig03}  we show the predictions of Model 5 with metallicity 
dependent yields
from K06 for both P and Fe. Even for these yields we
predict a too low solar ratio of $[P/Fe]_{\odot} \sim -0.40$ dex. 
In other words, although the trend of [P/Fe] in the Galactic disk is well reproduced, to have good agreement with the data the whole curve should be shifted upwards by 0.4 dex.

In the same Figure \ref{Fig03} we also show the predictions of Model 6
with the metallicity dependent yields from K06 for both P and Fe, but for
stars more massive than 20M$_{\odot}$ we adopted the yields for
hypernovae, instead of those for normal SNe II. 
The agreement with the observed solar ratio here is slightly better being $[P/Fe]_{\odot} \sim -0.37$ dex, although the problem of the too low predicted solar P abundance remains.

The agreement of Models 1,2,5 and 6 with the trend of the data and the 
disagreement with the observed solar abundances,  
suggests that the major producers of P in the
Universe should be core-collapse supernovae and that the theoretical
yields from these supernovae are underestimated by a factor of $\sim$ 3. 
 However, the results of Model 6 with yields from hypernovae differ 
substantially from those obtained with normal yields from SNe II at low 
metallicities.
In fact, the model with normal SNII yields predicts an overabundance 
of P relative to Fe at low metallicity of 
$\sim$+0.5 dex, whereas the model with hypernova yields predict a 
lower [P/Fe] ratio, 
namely an overabundance of $\sim $+ 0.2-0.3 dex. Clearly these
predictions await to be proven by future data in low metallicity halo
stars.



On the other hand, the
trend of abundances of [P/Fe] in the Galactic disk as predicted by  
Models 3 and 4,
adopting the metallicity dependent yields of P and Fe from massive
stars by WW95, is at variance with observations even if we normalize the model results to their predicted solar abundances.  
Actually, the predictions of Model 4 are very similar to
those of Timmes \& al. (1995): these authors adopted, in fact, the same
yields from massive stars for P and Fe as in Model 4, and showed the
predicted [P/Fe] vs. [Fe/H], but no data for P were available at that
time.

From our numerical simulations, we can also infer that the theoretical predictions from different authors 
for Fe are similar, whereas
for P the yields by K06 and by WW95 are significantly different.  We
should therefore conclude that the yields of P by K06 are better than
those of WW95, since the former can well reproduce the trend of [P/Fe]
vs. [Fe/H] and the P solar abundance by increasing them by a
factor of $\sim 3$.

\begin{figure}[htb]  
    \begin{center}  
    \includegraphics[width=0.48\textwidth]{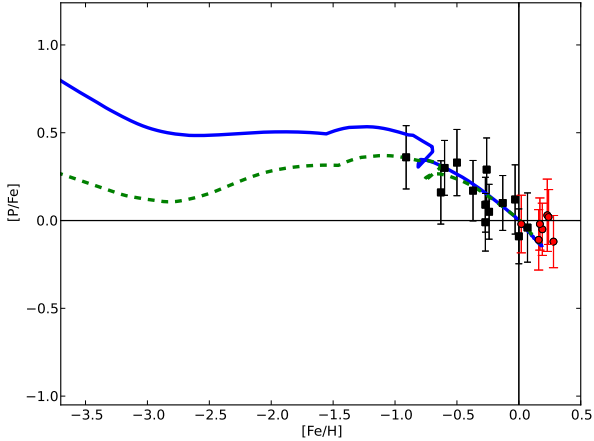}
    \caption{The predictions for [P/Fe] from Model 7 (blue solid line) and from Model 8
      (green dashed line). The P and Fe abundances in both models are normalized with the absolute solar values of Caffau et al. (2011b).
      Data as in Figure 1.}
    \label{Fig04} 
    \end{center}
    \end{figure}

For this reason and for the sake of having a more homogeneous set of
yields, we only show in the next figure the results obtained by including the 
K06
yields either with or without hypernovae.  In particular, 
Model 7 is based on normal massive stars yields whereas Model 8 is based on
hypernovae yields. In these models, shown in Fig. \ref{Fig04}, we apply
the necessary factor in the production of P in massive stars to
reproduce the observed solar abundance of P. In fact, as one can see in Figure 4, the predicted curve passes exactly by zero and this is because our model can reproduce, at the same time, the solar abundance of Fe and that of P.
 It is worth noting that
adopting the hypernova yields which are larger for P than those for normal
core-collapse SNe, the yields need to be increased by a factor
of $\sim$ 2.75, instead of $\sim$ 3. With the corrected yields and the observed solar abundance, the predicted [P/Fe] for [Fe/H]$<$ -1.0 dex is on average +0.2-0.3 dex.

\section{Conclusions}

In this paper we have compared model predictions adopting different
sets of yields with recent data on the abundance of P in the Galactic
disk. Our conclusions can be summarized as follows:

\begin{itemize}

\item The observed fall of the abundance of [P/Fe] in Galactic disk 
stars suggests that 
P is mainly produced by core-collapse SNe with a small contribution from SNe Ia.
Otherwise, if SNe Ia were important the  [P/Fe] ratio would remain fairly 
constant.

\item The metallicity dependent yields of P from massive stars of K06
  together with the P yields from SNe Ia from Iwamoto et al. (1999)
  well reproduce the data if the yields from massive stars 
  are increased by a factor of $\sim$ 3. 
  This suggests that the yields of P available in the literature
  are underestimated.
  Both the neutron rich isotopes of $^{29,30}$Si and $^{31}$P, which
  derives from neutron capture on the two Si isotopes, are produced in
  the oxygen and neon burning shells in massive stars
  (WW95). Therefore, in order to have a larger P production, the O and
  Ne shell- burnings as well as the neutron capture on the
  $^{29,30}$Si isotopes and the destruction of $^{31}$P by (p, $\alpha$) 
 reactions should be revised.

\item We predict also the
  behaviour of [P/Fe] in halo stars and we suggest that it should show a
  plateau between [Fe/H] =-1.0 and -3.0 dex corresponding to [P/Fe]
  $\sim$ +0.5 dex if yields (corrected) from normal SNe II are adopted, 
and to $\sim$+0.2dex if hypernova yields (corrected) are adopted. In order 
to distinguish between these two cases we await for observations of the P abundance in halo stars. 

\end{itemize}


\end{document}